\documentclass[prb,aps,twocolumn,superscriptaddress,showpacs,footinbib]{revtex4-2}
\usepackage[colorlinks=true,linkcolor=black, citecolor=blue, urlcolor=blue,
    unicode=true]{hyperref}

\usepackage{graphicx,amsmath,braket,tikz,amssymb,footmisc,lipsum,array,tabularx,float,comment}

\usepackage{cleveref}

\usepackage[utf8]{inputenc}
\usepackage{color}

\usepackage{subfigure}

\crefname{equation}{}{}
\Crefname{equation}{}{}
\crefrangelabelformat{equation}{#3#1#4--#5#2#6}

\crefmultiformat{equation}{#2#1#3}{, #2#1#3}{#2#1#3}{#2#1#3}
\Crefmultiformat{equation}{#2#1#3}{, #2#1#3}{#2#1#3}{#2#1#3}

\allowdisplaybreaks

\newcommand{\zwo}{ZrW$_{2}$O$_{8}$~}

\begin{document}


\title{Low energy phonons in single crystal ZrW$_{2}$O$_{8}$}

\author{R. A. Ewings}
\email[]{russell.ewings@stfc.ac.uk}
\affiliation{ISIS Pulsed Neutron and Muon Source, STFC Rutherford Appleton Laboratory, Harwell Campus, Didcot, Oxon, OX11 0QX, United Kingdom}

\author{A. I. Duff}
\affiliation{Scientific Computing Department, STFC Daresbury Laboratory, Warrington, WA4 4AD, United Kingdom}

\author{K. Refson}
\affiliation{ISIS Pulsed Neutron and Muon Source, STFC Rutherford Appleton Laboratory, Harwell Campus, Didcot, Oxon, OX11 0QX, United Kingdom}

\author{T. G. Perring}
\affiliation{ISIS Pulsed Neutron and Muon Source, STFC Rutherford Appleton Laboratory, Harwell Campus, Didcot, Oxon, OX11 0QX, United Kingdom}

\author{J. Ollivier}
\affiliation{Institut Laue-Langevin, 71 Avenue des Martyrs, CS 20156, 38042 Grenoble Cedex 9, France}


\date{\today}

\begin{abstract}

ZrW$_{2}$O$_{8}$ is the prototypical example of a material exhibiting negative thermal expansion (NTE). It is now widely accepted that in ZrW$_{2}$O$_{8}$, and in many other framework materials exhibiting NTE, a collection of low energy phonon modes, as opposed to just one or two, are responsible for the anomalous thermal properties. However, quantitative verification and analysis of the density functional theory (DFT) calculations which underpin this proposal are still lacking. In particular, probing the low energy phonons directly throughout reciprocal space using inelastic neutron scattering, as opposed to other techniques which only probe the Brillouin zone center, is technically challenging and hence rarely done. Here we report inelastic neutron scattering measurements in a large number of Brillouin zones over a 400\,K temperature range. We find excellent agreement between DFT calculations and experimental data at low temperature. However, the shifts in phonon modes predicted by DFT due to the reduction in lattice parameter (warming) are not observed. This is most likely due to counteractive anharmonic effects, which we verified using finite temperature molecular dynamics (MD) calculations. Notwithstanding, both DFT and MD results are consistent with NTE in \zwo arising from the tension effect, and by extension this explanation is supported by the neutron scattering results.
\end{abstract}

\maketitle

\section{Introduction}\label{sec:intro}

\begin{figure}[!h]
\centering
    \includegraphics[scale=0.44]{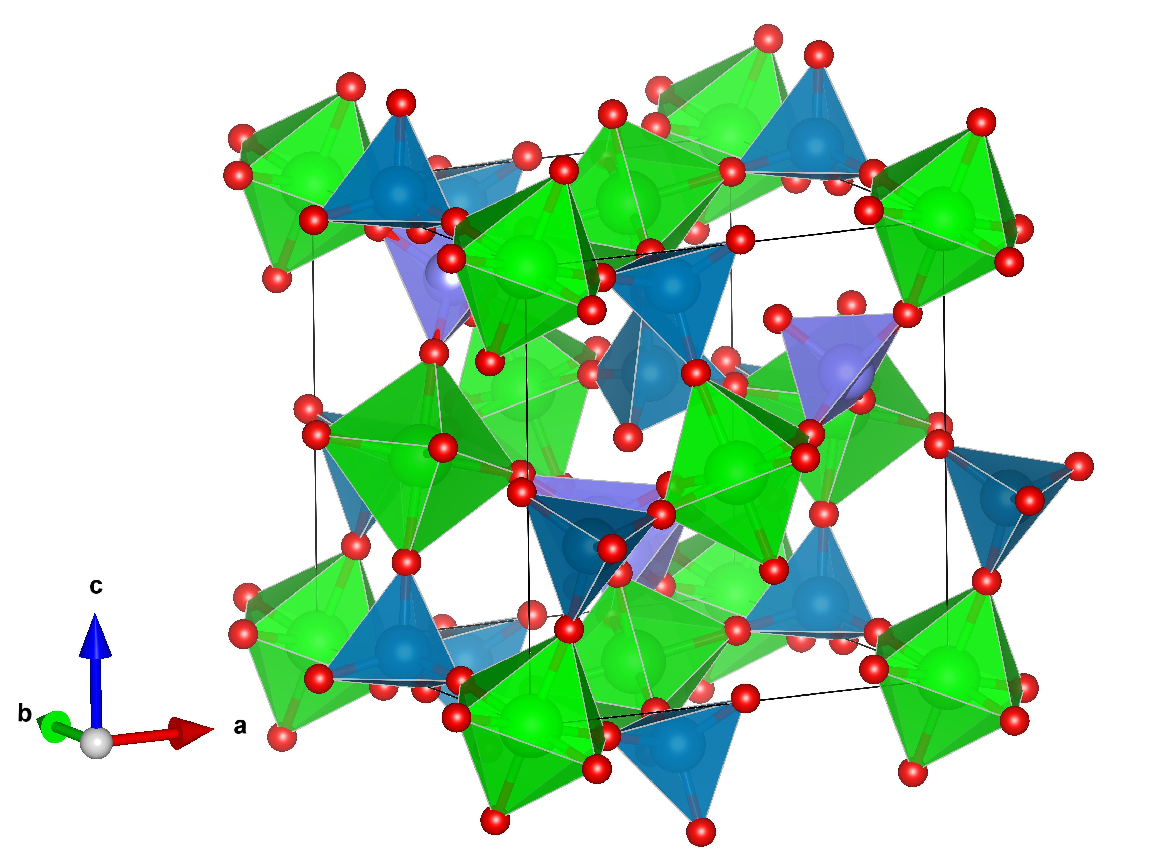}
    \caption{The crystal structure of \zwo (space group $P2_{1}3$) with ZrO$_{6}$ octahedra shaded green and WO$_{4}$ tetrahedra shaded purple and blue, corresponding to tetrahedra centred on the crystallographically distinct W1 and W2 sites respectively. Note that each WO$_{4}$ tetrahedron contains one unbonded oxygen ion on its vertex, pointing along a local $\langle$1\,1\,1$\rangle$-direction. This figure was created using VESTA \cite{VESTA}.}
    \label{f:structure}
\end{figure}

Negative thermal expansion (NTE) is when a material contracts rather than expands on warming. This effect can be uniform, i.e. along all crystallographic axes \cite{DAPIAGGI2003231,Greve-2010-ScF3,Lock-2010-MOF5}, or non-uniform \cite{Goodwin-Ag3CoCN6-uniaxial,Sartbaeva_2004,Grobler-2013-uniaxial}. It may also exist over a wide or a more limited range of temperatures \cite{KimFultz-Si-2018}. \zwo is in many respects the canonical example, with its NTE having been re-discovered in the 1990s \cite{ZWO-NTE-discovery,Mary1996,Evans-Science-1997}. It is particularly notable that its NTE behavior is uniform, extant over a wide temperature range, from $\sim 10$\,K to $\sim 1000$\,K, and its coefficient of thermal expansion is large and negative ($\sim 10^{-5}\rm{K}^{-1}$) (see review articles ref. \onlinecite{Dove2016} and ref. \onlinecite{MITTAL2018} and references therein). In common with many other materials known to exhibit NTE, \zwo has a framework structure, i.e. it contains many ions that form coordinated polyhedra, most of which share corners, with large pores or spaces between the polyhedra (see fig. \ref{f:structure}. It has also been found that intercalation of water into these pores can also result in significant anomalous lattice expansion or contraction \cite{Baise2018}.

As a material is warmed, phonon modes with gradually increasing energy are populated. If the potentials for these modes are asymmetric as a function of displacement, then thermal expansion will arise \cite{Dove-book-old,Dove-book-new}. Studies of the phonons in NTE materials often involve measurement of the phonon density of states (PDOS) using either Raman spectroscopy, inelastic x-ray scattering, or inelastic neutron scattering measurements of powder samples \cite{Ravindran-Raman-2003,Oishi_2017,Mittal-2009-ApplPhysLett-PDOS,CHAPMAN2006,GUPTA2012-Ag2O,Mittal-2007-Cu2O-Ag2O,Li-2011-ScF3}, as this is relatively straightforward to do and can be compared directly to quasi-harmonic lattice dynamics calculations such as those from density functional theory (DFT). \zwo is no exception, with the phonon density of states having been measured as a function of temperature soon after the seminal works on the NTE behavior \cite{Ernst1998}. It was found that the low energy peaks in the PDOS shift to higher energy on warming. Qualitatively, this is the opposite behavior to normal (positive thermal expansion) in which increased bond lengths on warming typically result in weaker interatomic potentials and hence a reduction in the energy of phonon modes.

In terms of interpretation of the PDOS measurements, and NTE more generally, early theories concerned the role of rigid unit modes (RUMs), in which  Zr--O and W--O polyhedra remain undistorted but the whole structure flexes considerably along their shared vertices. As the amplitude of such modes increases, the root-mean-square distance between metal centers decreases, even though individual metal--oxygen bonds do not change in length \cite{Dove2016}. Such modes would typically be found at the lower energy end of the phonon spectrum. However, various measurements have indicated that the picture may be somewhat more complex than this. In particular, it has been found from EXAFS \cite{Bridges2014,Vila-FChem-Calc-2018,Vila-FChem-EXAFS-2018} that all of the polyhedra in the structure distort somewhat on warming, although the WO$_{4}$ tetrahedra are noticeably more rigid than the ZrO$_{6}$ octahedra. Notwithstanding, significant flex occurs along some of the Zr--O--W bonds that comprise the corner-sharing polyhedra. In this context, it would be appealing to search for a particular phonon mode that facilitates this distortion. However, numerous computational studies of the lattice dynamics in \zwo \cite{Rimmer2015,Sanson2014,Gupta2013,FigueiredoPRB2007} have demonstrated that there is no such single mode responsible for NTE. Rather, a large number of low energy phonon modes distributed throughout the Brillouin zone have a character that leads to NTE.

The direct measurements of phonons in \zwo (and many other NTE materials) to date have had a number of limitations. Raman spectroscopy can only measure the Brillouin zone center ($\mathbf{q}=0$) Raman-active modes and thus does not provide a comprehensive picture, although the energy resolution available with this technique is usually superior to that available with neutron or x-ray spectroscopy. Neutron (or x-ray) scattering PDOS measurements are by their very nature a spherical average over reciprocal space of all possible phonon modes at a particular energy, so there is no direct sensitivity to the Q-space origin of shifts in energy with temperature. This can be ameliorated somewhat by interpreting these measurements with DFT calculations. However, this approach is challenging in systems containing many atoms, and hence many phonon modes which may be more closely spaced than the energy resolution of the neutron spectrometer used for the measurements. In contrast, a measurement of a single crystal sample would allow individual modes to be isolated via measurements performed in many different Brillouin zones (see appendices for a more detailed discussion). This would then provide a more stringent test of the DFT.

The motivation of the present study is therefore to develop a more comprehensive understanding of the low energy phonons in \zwo, by measuring in more Brillouin zones using inelastic neutron scattering with good energy resolution, to identify if any particular individual or groups of phonon modes are especially sensitive to temperature. Such measurements are sensitive to both the eigenvalues and eigenvectors extracted from \emph{ab-initio} calculations, and hence pose a stringent test of the range of their validity, most notably for the low energy phonons thought to be important for NTE.

\section{Methods}

Time of flight neutron scattering experiments were performed on the IN5 spectrometer at the Institut Laue-Langevin \cite{Ollivier-IN5}. Data were collected with incident neutron energies $E_{i} = 8.24$\,meV and $E_{i}=10.43$\,meV, at temperatures in the range 2\,K to 400\,K. Density Functional Theory (DFT) calculations were performed using the plane-wave pseudopotential method as implemented in the CASTEP package \cite{ClarkCASTEP}. Calculations were performed for a series of different fixed lattice parameters close to the ground state parameters determined from a minimization of the overall forces and energies. Phonon dispersions were calculated using Density Functional Perturbation Theory (DFPT). It is important to note that in DFT the lattice dynamics are calculated at zero temperature in the quasi-harmonic approximation. When comparing the computed dispersion with that measured experimentally with neutrons, an empirical scale factor was applied to the eigenvalues (phonon energies) but not to the eigenvectors (structure factors).

Calculations of the finite-temperature behavior of the phonon dispersion at a few high-symmetry wavevectors were performed using machine learned potentials and molecular dynamics (MD). The potentials were computed using the automated potential development (APD) workflow \cite{duff2023automated}. These potentials were then used for MD simulations, using the LAMMPS software package \cite{plimpton2007lammps}.

Further details of the inelastic neutron scattering experimental methods can be found in the appendices, as can the detailed schemes used for the density functional theory and molecular dynamics calculations.

\section{Results}\label{sec:results}

\subsection{INS data}

\begin{figure*}
\centering
    \includegraphics[scale=0.85]{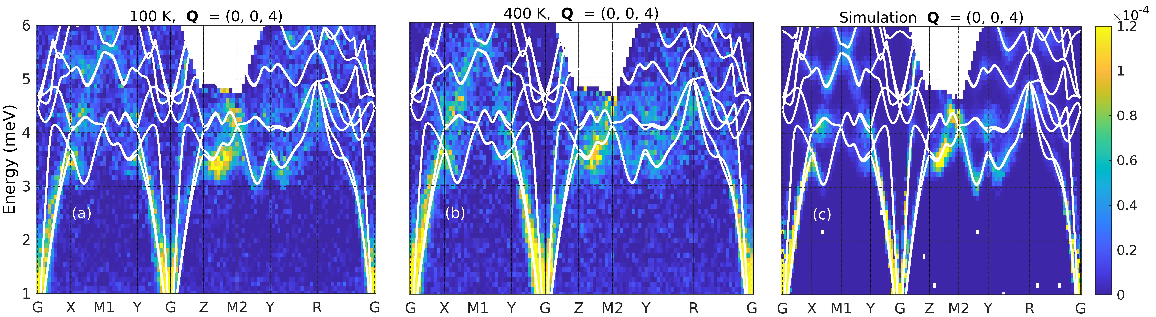}
    \caption{Slices through the $\mathbf{Q}$--Energy plane, following a defined path through along high-symmetry reciprocal space directions. The labeling convention for these high-symmetry positions is $\rm{G}\equiv\Gamma =(0,0,0)$, $\rm{X}=(\frac{1}{2},0,0)$, $\rm{M1}= (\frac{1}{2},\frac{1}{2},0)$, $\rm{Y} = (0,\frac{1}{2},0)$, $\rm{M2}= (0,\frac{1}{2},\frac{1}{2})$, $\rm{Z}=(0,0,\frac{1}{2})$, and $\rm{R}=(\frac{1}{2},\frac{1}{2},\frac{1}{2})$. The data and simulations shown in this figure were centered on the $\mathbf{Q}=(0,0,4)$ Brillouin zone. The neutron scattering structure factor (intensity) is indicated by the false colormap. Panel (a) shows data collected at 100\,K, panel (b) shows data collected at 400\,K, and panel (c) shows simulations of the neutron cross-section computed from the DFT calculations with a lattice parameter of $9.215$\,\AA. The data in panels (a) and (b) have had a constant background subtracted and have then been corrected for the Bose-Einstein population factor so that their color scales (which are in arbitrary but consistent units) can be compared directly. The color scale for the simulation in panel (c) is scaled so that the 5\% most intense data points match their equivalents in panel (a). The white lines in all three figures correspond to the dispersion relation calculated using DFT with a lattice parameter of $9.215$\,\AA.}
    \label{f:multispag004}
\end{figure*}

\begin{figure*}
\centering
    \includegraphics[scale=0.85]{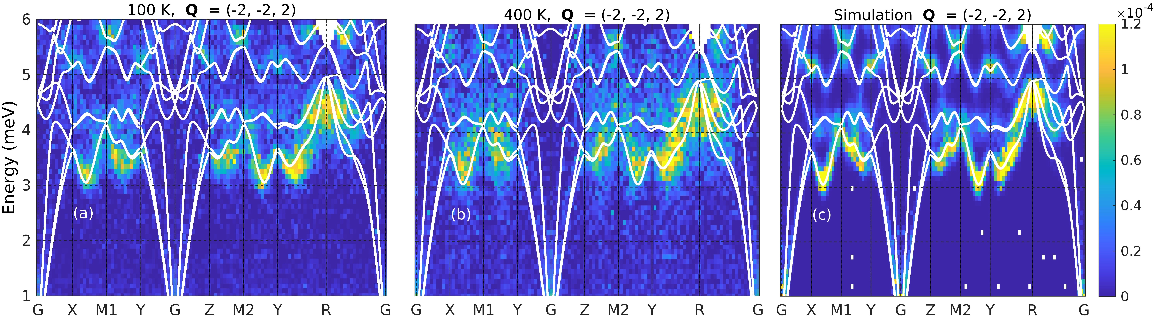}
    \caption{Slices through the $\mathbf{Q}$--Energy plane, following the same scheme as fig. \ref{f:multispag004}. The data here were centered on the $\mathbf{Q}=(-2,-2,2)$ Brillouin zone. Notice that the change in direction of $\mathbf{Q}$ results in the highest intensity being peaked at different high symmetry positions than for the data shown in fig. \ref{f:multispag004}.}
    \label{f:multispag222}
\end{figure*}

\begin{figure*}
\centering
    \includegraphics[scale=0.85]{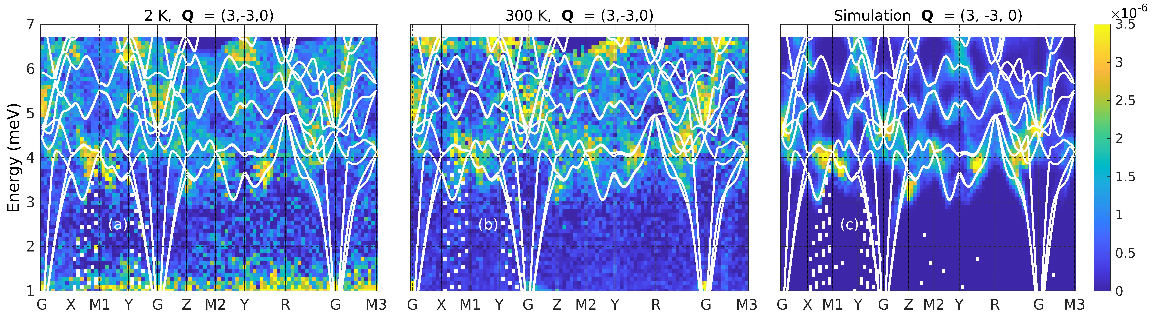}
    \caption{Slices through the $\mathbf{Q}$--Energy plane, following the same scheme as fig. \ref{f:multispag004}, with the exception of the extra reciprocal lattice point on the right hand side that is labeled M3 and refers to ($\frac{1}{2},-\frac{1}{2},0)$. The data here were collected in a separate experiment and are centered on the $\mathbf{Q}=(3,-3,0)$ Brillouin zone. Data shown were collected at 2\,K and 300\,K, which is different to figs. \ref{f:multispag004} and \ref{f:multispag222}. The different experiments also have a different (but mutually consistent) set of arbitrary units for the intensity scale. These data are notable by having intensity peaked near the G point, unlike the other two datasets shown, and extending to slightly higher energy.}
    \label{f:multispag330}
\end{figure*}

Figure \ref{f:multispag004} shows slices through the 4-dimensional datasets recording $S(\mathbf{Q},\omega)$, in the $\mathbf{Q}$--Energy plane \footnote{By convention energy is denoted by frequency $\omega$ in the expression of the neutron structure factor.}, following a path through reciprocal space traversing a set of high-symmetry points. Note that although the points labeled as X, Y, and Z; and as M1, M2 and M3 are equivalent by symmetry, there are paths between them that are inequivalent by symmetry because the space group, $P2_{1}3$, is non-centrosymmetric. Data are presented centered at $\mathbf{G} = (0,0,4)$ in fig. \ref{f:multispag004} and $\mathbf{G} = (-2,-2,2)$ in fig. \ref{f:multispag222} for both $T = 100$\,K and $T = 400$\,K, and at $\mathbf{G} = (3,-3,0)$ in fig. \ref{f:multispag330} for $T= 2$\,K and $T = 300$\,K. The DFT-calculated dispersion relation with $a = 9.215$\,\AA, with the energy scale reduced by 5\%, is overplotted. 
The calculated structure factor is also shown in panel (c) of all three figures. This set of figures illustrate many of the salient points of this work. It is clear that the dispersion and cross-section calculated using DFT provide an extremely good level of agreement with the neutron measurements, with significant overlap at most wavevectors, indicating that both the eigenvalues and eigenvectors from the DFT are accurate.

\begin{figure}[!h]
\centering
    \includegraphics[scale=0.47]{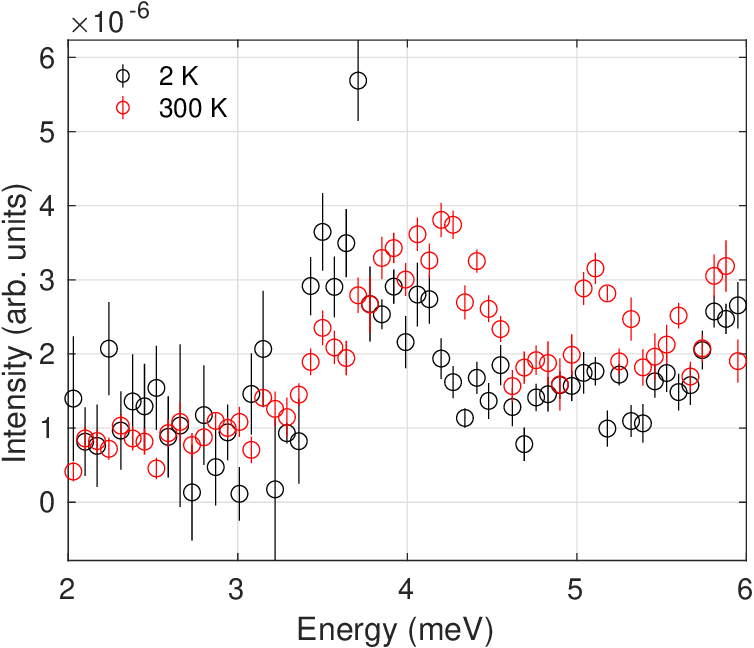}
    \caption{Cut at the mid-point of the Y--R trajectory of fig. \ref{f:multispag330}(a) and (b), i.e. at $\mathbf{Q} = (3.25,-2.5,-0.25)$, showing hardening of the first phonon peak from $\sim 3.5$\,meV to $\sim 4$\,meV.}
    \label{f:YRcut_Tdiff}
\end{figure}

\begin{figure}[!h]
\centering
    \includegraphics[scale=0.47]{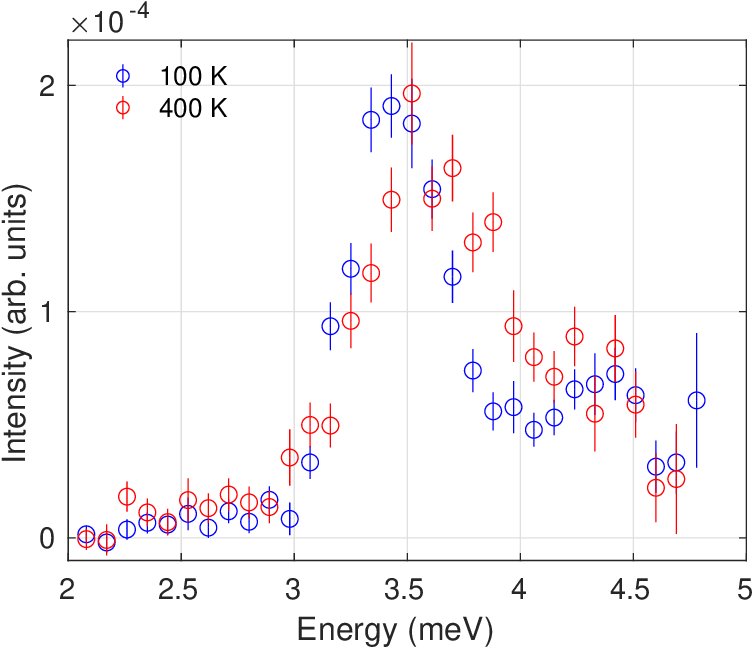}
    \caption{Cut at the mid-point of the Z--M2 trajectory of fig. \ref{f:multispag004}(a) and (b), i.e. at $\mathbf{Q} = (0,0.25,4.5)$, showing a slight hardening of the first phonon peak from $\sim 3.4$\,meV to $\sim 3.5$\,meV.}
    \label{f:ZM2cut_Tdiff}
\end{figure}

At most wavevectors a change in temperature of $\sim 300$\,K results in very little change in the measured dispersion and $S(\mathbf{Q},\omega)$. A broadening of the signal from most of the phonons modes, indicative of anharmonicity at elevated temperatures, is observed. One notable exception is that on warming from 2\,K to 300\,K a significant mode hardening of around 0.5\,meV is seen on the mode between the Y and R points in fig. \ref{f:multispag330}. To show this in more detail a cut at the mid-point, $\mathbf{Q} = (3.25,-2.5,-0.25)$, is shown in fig. \ref{f:YRcut_Tdiff}. Modes along similar trajectories in the reduced Brillouin zone in the other datasets (figs. \ref{f:multispag004} and \ref{f:multispag222}) do not show such a dramatic shift. However, we can see from careful inspection of the DFT-calculated dispersion at the Y point that what appears to be a single mode is actually two different ones with only a small separation in energy. The fact that the two modes appear to behave differently with temperature is a good illustration of the power of measuring in multiple Brillouin zones - with only the data centered on $(-2,-2,2)$ and $(0,0,4)$ this effect would not have been visible. A similar, although less dramatic, mode hardening occurs at the mid-point of the Z--M2 trajectory (equivalent by symmetry to Y--M1), which is visible in both fig. \ref{f:multispag004} and \ref{f:multispag222}, and is shown as a cut in fig. \ref{f:ZM2cut_Tdiff}.

Despite the overall excellent agreement between the DFT calculations and the INS measurements shown in figs. \ref{f:multispag004}, \ref{f:multispag222} and \ref{f:multispag330}, there is some disagreement. An example can be seen by comparing the calculated dispersion and cross-section shown in panel (c) with the data in panels (a) and (b) of all three figures. For example, we can see that for the lowest energy mode on the path M1 -- Y (and M2 -- Z, which is equivalent by symmetry) the calculated energy is rather higher than the measurement at 100\,K, whereas the other low energy ($< 4$\,meV) modes along other paths agree rather closely at this temperature.

\begin{figure}
\centering
    \includegraphics[scale=0.65]{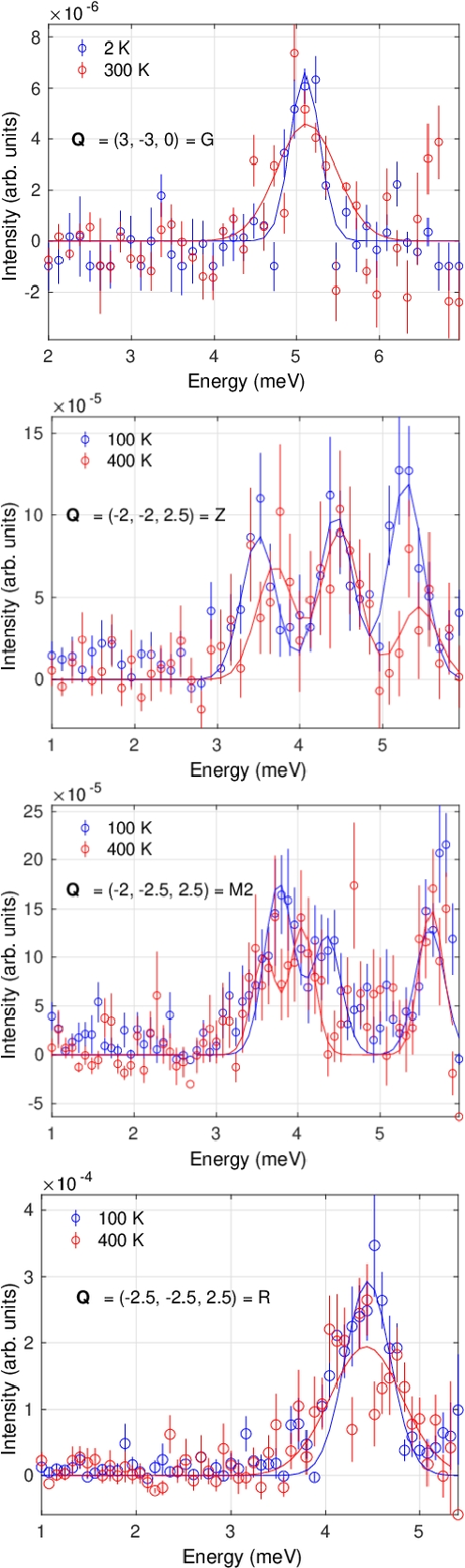}
    \caption{Cuts along the energy axis at four high symmetry positions in reciprocal space, centered on different Brillouin zones in order to maximize the cross-section. Panel (a) shows a cut at G, panel (b) a cut at Z, panel (c) a cut at M2, and panel (d) a cut at R. Data at lower temperatures (2\,K in panel (a), 100\,K in panels (b)--(d)) are shown in blue whereas data at higher temperatures (300\,K in panel (a), 400\,K in panels (b)--(d)) are shown in red. For all cuts an estimated constant background has been subtracted, followed by correction for the Bose-Einstein population factor in order to allow easier comparison of data taken at different temperatures. Solid lines correspond to fits to Gaussians (the number of which is set by the number expected from the DFT calculations), to indicate the peak positions.}
    \label{f:multicuts}
\end{figure}

To allow a more quantitative analysis we show constant $\mathbf{Q}$ cuts along the energy axis at various high symmetry wavevectors, G, X, M and R, at different temperatures in fig. \ref{f:multicuts}. In all cases the integration range along the orthogonal $q$-axes is 0.04 reciprocal lattice units. This finite integration range, necessary when plotting data from time-of-flight INS experiments, obviously means that for modes that are highly dispersive near the wavevector of interest the width of peaks may be broader than the nominal instrumental energy resolution. This is most apparent for the single peak in the cut at R -- referring to fig. \ref{f:multispag222} one can see that the dispersion is very steep until quite close to the R point.

\subsection{DFT calculations}\label{ss:DTFcalcs}
The DFT calculations plotted in figures. \ref{f:multispag004} -- \ref{f:multispag330} were performed at a lattice parameter of 9.215\AA, and we then reduced all calculated phonon energies by 5\% to give best agreement with the INS measurements \footnote{This was done by doing a least squares fit of the neutron spectra with the DFT-calculated spectra, varying only the scale factor to give the best agreement}. 

\begin{figure}[h]
\centering
    \includegraphics[scale=0.42]{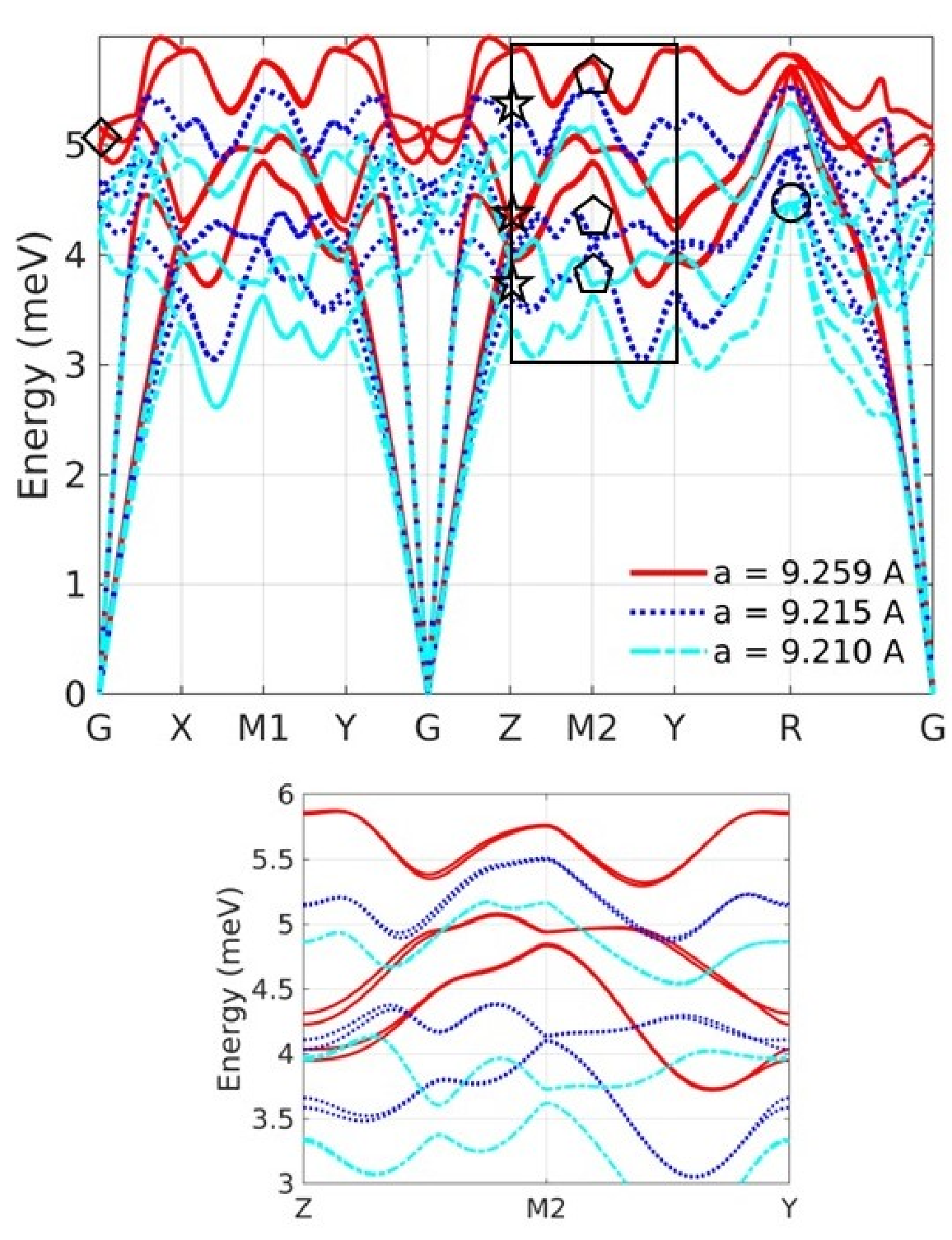}
    \caption{Comparison of dispersion from DFT calculations with different lattice parameters. The reciprocal space path is the same a described in fig. \ref{f:multispag004}. The solid red curve is for calculations with lattice parameter $a = 9.259$\,\AA\, (zero stress, optimized lattice parameter), the dotted blue curve is for calculations with $a = 9.215$\,\AA, and the dash-dotted cyan curve is for calculations with $a = 9.215$\,\AA. The inset shows a zoomed in section of the dispersion between 3 and 6\,meV. The symbols at high symmetry wavevectors correspond to the fitted peak positions for the lower temperature datasets shown in fig. \ref{f:multicuts}, with diamonds for the Gamma - point, stars for the Z - point, pentagons for the M2 - point and a circle at the R - point.}
    \label{f:disp_latt}
\end{figure}

We show in fig. \ref{f:disp_latt} the effect of the lattice parameter on the computed phonon spectra below 6\,meV (the energy range covered in our INS experiments). As would be expected from the Gr\"uneisen parameters reported in the literature, the low energy phonons are mostly found to harden with increasing lattice parameter. However, the scaling is not uniform, as can be seen in the inset which shows the trajectory along Z -- M2 -- Y. For instance, the two lowest energy modes for $a = 9.210$\,\AA\, and $a=9.215$\,\AA\, are relatively flat in energy, whereas the equivalent modes for $a = 9.259$\,\AA\, are much more dispersive. Furthermore, one can see that at the M2 point the highest energy mode decreases by $\sim 0.2$\,meV on compression of the lattice from $a = 9.259$\,\AA\, to $a=9.215$\,\AA\, ($\Delta a = -0.044$\,\AA) and decreases by even more, $\sim 0.3$\,meV, for further compression $\Delta a = -0.005$\,\AA. As noted above, we compared the INS data with the calculations for the unit cell with $a = 9.215$\,\AA\, rather than $a = 9.259$\,\AA, which is the zero-stress optimized unit cell dimension in the DFT, or the cell with $a = 9.210$\,\AA. Globally rescaling the phonon energies in either of the latter two sets of calculations resulted in worse agreement with the neutron data.

As noted, a small compression of the lattice from $a = 9.215$\,\AA\, to $a = 9.210$\,\AA\, yields a rather large change in mode energies, suggesting that around these lattice parameters the calculations are on the edge of dynamical stability. Indeed, when the lattice was compressed further in the DFT calculations (not shown), a significant number of soft modes appeared, consistent with this interpretation. With further compression negative energy modes appear, indicating a change in the ground state structure. Given this non-linear behavior, to make a quantitative prediction for the effect on the phonon energies of temperature-induced lattice compression in this quasi-harmonic approximation, it makes most sense to compare the calculations with $a = 9.259$\,\AA\, and $a=9.215$\,\AA, rather than those with $a = 9.215$\,\AA\, and $a = 9.210$\,\AA. The relative difference in the former two lattice parameters is 0.48\%, which is quite close to the experimentally observed change in lattice parameter on warming from 100\,K to 400\,K (0.37\%). Thus, one can take the changes in the phonon dispersion between these two calculations as representative of the changes in the phonon dispersion over the experimentally probed temperature range, based on the Gr\"uneisen model alone. The results of this procedure at various high symmetry wavevectors are presented in table \ref{tab:peakpos}.

\begin{table*}[ht]
    \caption{Comparison of the position of the lowest energy peak in the measured data at 100\,K and 400\,K, and the expected change in peak position based on DFT alone (i.e. quasi-harmonic approximation), and the computed change in peak position from the work of Gupta \emph{et al} \cite{Gupta2013}, decomposed into the contributions from implicit and explicit anharmonicity, from a phenomenological model.}\label{tab:peakpos}
\begin{ruledtabular}
\begin{tabular}{p{1cm}p{2.7cm}p{2.7cm}p{2.7cm}p{4cm}}
 Wavevector & Lowest energy peak at 100\,K (meV) & Lowest energy peak at 400\,K (meV) & DFT shift (meV) [this work] & Calculated shift [implicit + explicit] (meV) \cite{Gupta2013} \\
\hline
  G & 5.10(3) & 5.12(5)  & -0.55 & -0.07 [-0.22 + 0.15]  \\  
  Z & 3.50(5) & 3.70(10) & -0.40 & 0.08 [-0.14 + 0.22]  \\  
  M & 3.77(4) & 3.58(7)  & -0.78 & 0.06 [-0.36 + 0.42]  \\  
  R & 4.45(2) & 4.43(5)  & -0.80 & -0.77 [-0.39 - 0.38]\\
\end{tabular}
\end{ruledtabular}
\end{table*}


The DFT calculations presented so far correspond to the lattice contracting or expanding under pressure / stress, rather than warming / cooling. In reality, the warming will result in an increased amplitude of atomic vibrations, which is not accounted for in the quasi-harmonic approximation. Anharmonic effects may become more significant at higher temperatures. This point has been discussed extensively by Mittal \emph{et al} \cite{Gupta2013,MITTAL2018}, who have calculated these effects for a few individual modes at particular wavevectors in \zwo. Some attempt has been made to quantify the effect more broadly using Raman spectroscopy data \cite{Ravindran-Raman-2003,Oishi_2017}, which has indicated that as well as having large negative Gr\"uneisen parameters, the lowest energy modes also have the largest anharmonicity.

The effects of temperature on phonons can be decomposed into two parts. On the one hand a change in lattice parameter with temperature often gives rise to a changed bond length between atoms hence a change in phonon energy. This is sometimes referred to as implicit anharmonicity. In conventional materials, increased temperature gives longer and weaker bonds, and hence softened phonons. On the other hand, as temperature increases phonon-phonon interactions become increasingly important, sometimes also referred to as explicit anharmonicity, and renormalize the phonon energies. This often yields phonon modes that harden slightly with increasing temperature \cite{Dove-book-new}. However, in practice in conventional materials this effect is often much smaller than the softening due to thermal expansion and so is masked.

In table \ref{tab:peakpos} we compare the positions of peaks in the INS signal at high symmetry wavevectors as a function of temperature, with the expected shift due to a change in lattice parameter alone from the DFT. Additionally, we compare these results to phenomenological calculations by Gupta \emph{et al} \cite{Gupta2013} that decompose the shift into components arising from implicit and explicit anharmonicity. With the exception of the calculation at the R--point, the changes in peak position from the latter are much smaller than those predicted from the DFT alone, and are closer in magnitude to the effects seen in the INS measurements.

\subsection{Machine learned potentials and molecular dynamics calculations}\label{ss:MLMD}

An alternative way of analyzing the phonons, that explicitly includes finite temperature effects, is to use molecular dynamics (MD) calculations. The approach we took was to use the APD framework for machine-learned interatomic potentials \cite{duff2023automated}, an approach for \zwo closely related to that used in ref. \onlinecite{he2022origin}. A detailed technical description of the methodology is provided in the appendices. The resulting interatomic potentials can be used to interrogate material properties at finite temperatures. For convenience, we used the PBESol functional for the underpinning DFT that was used to generate the training data for the potentials. The potentials from this were optimized after 4 active learning cycles. As a check on results, the temperature-dependent lattice parameter was computed. A coefficient of linear thermal expansion of $-6.04 \times 10^{-6}$\,$\rm{K}^{-1}$ was found. This is a slight underestimate compared to the experimental value of $-9.37 \times 10^{-6}$\,$\rm{K}^{-1}$, but it is crucially the correct sign and is similar in magnitude to that found by He \emph{et al} \cite{he2022origin}. The phonon dispersion obtained from the trained potentials was also verified against the zero temperature DFT dispersions.

In order to compare the MD results to the INS data, dynamical structure factor calculations were performed with the lattice parameters fixed to the time-averaged values at each temperature. Gaussians were then fitted to the peaks in energy scans up to 6\,meV at several high symmetry wavevectors to quantify the temperature-dependent shifts in these modes. The results are shown in fig. \ref{f:MD_phon_tdep}, with the energy scans shown in the appendices figs. \ref{f:MD_Escan_100} -- \ref{f:MD_Escan400}. The majority of the modes harden slightly on warming, with the exception of the highest energy one at the M point ($\mathbf{q} = (0.5,0.5,0)$) which softens slightly. We note that between 100\,K and 400\,K the calculated magnitude of mode hardening of $\sim 0.1$\,meV is similar in magnitude to what is seen in our INS measurements. But between 0\,K and 100\,K a much greater mode hardening is predicted, notably with the modes at $\mathbf{q} = (0,0.5,0)$ and $\mathbf{q}=(0.25,0.5,0)$ hardening on warming by $>0.5$\,meV, which is not observed in the measurements. This may, for example, be due to the precise choice of functional used for the DFT training data on which the interatomic potentials were based. We note that the DFT calculations presented in sec. \ref{ss:DTFcalcs} had to be fine-tuned for best agreement with the data, whereas the DFT training data for the MD calculations were not adjusted in this way.

\begin{figure}[h]
\centering
    \includegraphics[scale=0.65]{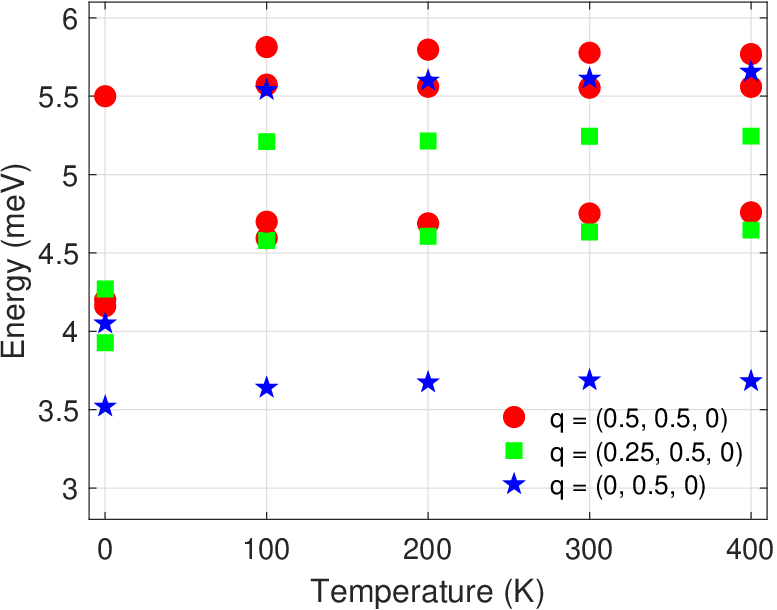}
    \caption{Temperature dependence of phonon modes below 6\,meV computed using MD with machine-learned potentials. Red circles indicate the modes at $\mathbf{q} = (0.5,0.5,0)$ (M1 in the notation of fig. \ref{f:multispag004}), green squares the modes at $\mathbf{q} = (0.25,0.5,0)$, and blue circles the modes at $\mathbf{q} = (0,0.5,0)$ (Y in the notation of fig. \ref{f:multispag004}).}
    \label{f:MD_phon_tdep}
\end{figure}

\section{Discussion and Conclusions}\label{sec:discussion}

From the results and analysis presented, we can see that there is very good agreement between the phonon dispersion relations and structure factors measured in our inelastic neutron scattering experiments, and those computed from DFT in the zero-temperature quasi-harmonic approximation, albeit with a compressed lattice compared to the optimized value. The agreement is good throughout reciprocal space, which is important in the context of the view that there is no single phonon mode responsible for NTE in ZrW$_{2}$O$_{8}$, but many. This therefore supports the usefulness of the DFT results in interpreting the physics of NTE in ZrW$_{2}$O$_{8}$. We do not observe any significant change in the neutron structure factors as a function of temperature, aside from some modes broadening on warming, so we can be confident that the phonon eigenvectors do not change significantly with temperature. The fundamental nature of the atomic motions thus do not change with temperature, which is consistent with DFT.

On the other hand, the DFT predicts softening of most phonon modes on compression of the lattice, and with NTE such a compression happens on warming. But we observe that the majority of phonons do not soften, indeed some harden by an amount ($\sim 0.1$\,meV) that is just discernible with our instrumental resolution. This is because DFT's quasi-harmonic calculations only capture the lattice compression on warming, but do not capture the full anharmonicity, which plays an increasing role with increasing temperature. Finite temperature effects can be included in MD simulations, which in our case used potentials generated through machine learning applied to DFT. Our MD simulations showed that the low energy modes at a few high-symmetry wavevectors do indeed harden slightly on warming. A larger hardening than what is observed was predicted between 0\,K and 100\,K, but as already noted this may be due to the specific choice of functional for the DFT training data for the generation of the interatomic potentials.

Given that the DFT calculations give good phonon eigenvectors when compared to experiment, and the MD calculations show the right behavior of the phonon eigenvalues as a function of temperature when compared to experiment, it is instructive to examine the underlying atomic motions that are common to both to help us understand NTE in \zwo. Such analysis has been done separately for the DFT \cite{Rimmer2015} and MD \cite{he2022origin} respectively. The former analysis compares the DFT to a set of simplified flexibility models, and concludes that NTE arises due to a tension effect involving nearly rigid Zr--O and W--O bonds, with low energy phonons likely involving quasi-rigid small whole-body rotations of the WO$_{4}$ tetrahedra. The analysis of the MD shows that with increasing temperature the Zr--O and W--O bond lengths indeed change very little, with a small positive expansion, but the W--O--Zr bond angle changes significantly, that could also be described as a tension effect. Our INS data are consistent with both sets of analysis, and both point towards the same basic mechanism of NTE.

\begin{acknowledgements}
We are grateful to D. Voneshen (ISIS) and R. Fair (STFC-SCD) for their advice and assistance on the use of Euphonic. We are indebted to M. Dove (QMUL) for his insightful and informative comments. Raw data are available via ref. \onlinecite{Data-DOI} and processed data via ref. \onlinecite{ProcessedData-DOI}. We thank the Institut Laue-Langevin for the allocation of beamtime for the experiments reported here. Computing resources were provided by the STFC Scientific Computing Department’s SCARF cluster. AID gratefully acknowledges funding from the Ada Lovelace Centre (ALC) at the Science and Technology Facilities Council (STFC) and from the Norwegian Research Council under grant "Neutron Scattering and Atomistic Simulations for a SUPERior Understanding of SUPERionic Conduction" (grant no. 344062).
\end{acknowledgements}

\newpage

\onecolumngrid

\section*{Appendix I: Methods}\label{appsec:methods}

\subsection*{A: Inelastic neutron scattering}\label{appsubsec:INS}

\zwo crystallizes in the $P2_{1}3$ cubic space group. The single-crystalline samples for these experiments were grown as described previously \cite{Kowach2000}. The sample used was composed of one crystal of mass 2.4\,g with a mosaic of $\sim0.3$ degrees, which is substantially less than the divergence of the incident neutron beam in the inelastic neutron scattering experiments described below, and hence unimportant for the experimental resolution. The lattice parameter was measured to be 9.176\,\AA\, at 2\,K, 9.159\,\AA\, at 100\,K, 9.1389\,\AA\, at 300\,K, and 9.125\,\AA\, at 400\,K during the alignment procedure for the inelastic scattering measurements described below. These values are consistent with those reported in the literature for powder samples \cite{Mary1996}.

The time of flight (ToF) neutron scattering experiments were performed on the IN5 spectrometer at the Institut Laue-Langevin \cite{Ollivier-IN5}. Data were collected with incident neutron energy $E_{i}=8.24$\,meV, with the final chopper spun at 16,000\,r.p.m. to give a resolution at the elastic line of 0.27\,meV (full-width half-maximum, FWHM). The sample was aligned so that the $(H,H,0)$ and $(0,0,L)$ reciprocal space directions were in the horizontal plane of the instrument. The data were collected by scanning the sample orientation about the vertical axis in 0.5 degree steps over a range of 161 degrees, starting from the $(1,1,0)$-axis perpendicular to the incident neutron beam. Data were collected at temperatures of 100\,K and 400\,K, with these temperatures chosen in order to increase the signal via the Bose-Einstein population factor while maintaining a large temperature difference between measurements, but avoiding the structural phase transition at $\sim 425$\,K \cite{Mary1996}. Additional data were collected with the sample aligned with the $(H,0,0)$ and $(0,K,0)$ directions in the horizontal plane of the spectrometer, with the sample orientation scanned over a range of 88 degrees in 1 degree steps. These data were collected with $E_{i}=10.43$\,meV and a final chopper speed of 16,000\,r.p.m. giving a FWHM energy resolution at the elastic line of 0.35\,meV. Data with this configuration were collected at temperatures of 2\,K and 300\,K.

The raw data were processed (units conversion, normalization of intensities, etc.) using the Mantid software package \cite{ARNOLD2014}. Further processing of the data, and all subsequent visualization and analysis, including convolution of calculated spectra with the instrumental resolution, were performed using the {\sc Horace} software package \cite{Ewings-Horace}. Calculation of the inelastic neutron scattering cross-section ($S(\mathbf{Q},\omega)$) from the DFT-calculated phonons was done using the Euphonic software package's {\sc Horace} plugin \cite{Euphonic-manual,Fair2022}.

\subsection*{B: Density functional theory}\label{subsec:DFT}

DFT calculations were performed using the plane-wave pseudopotential method as implemented in the CASTEP package \cite{ClarkCASTEP}. The Perdew-Burke-Ernzehof (PBE) exchange-correlation functional \cite{PBE-1996} was used with norm-conserving pseudopotentials of the RRKJ type \cite{Pseudopot-Ramer-1999} from the Rappe and Bennett library \cite{Opium-webpage}, a plane-wave cutoff energy of 750\,eV, and a $3 \times 3 \times 3$ Monkhorst-Pack \cite{Monkhorst-Pack-1976} grid for the Brilloun-Zone  of integration of electronic states. As also observed by Rimmer \emph{et al} \cite{Rimmer2015}, the equilibrium lattice parameter at zero pressure is 9.259\,\AA, slightly larger than the measured low temperature lattice constant of $\sim 9.18$\,\AA, typical of the PBE functional \cite{PBESol-2008}.

The phonon dispersion was calculated using Density Functional Perturbation Theory (DFPT) \cite{Refson-DFPT-2006}, with a $3 \times 3 \times 3$ Monkhorst-Pack grid of phonon wavevectors and Fourier interpolation of the resulting dynamical matrices, in order to sample the entire Brillouin zone. Additional calculations were performed for fixed unit cells of smaller dimension (9.23, 9.22, 9.215, 9.21 and 9.16\,\AA), at each of which the structure was relaxed and phonons calculated in the same way as for the fully relaxed structure. The calculated stress for each of these cell dimensions was 0.81, 1.13, 1.29, 1.45 and 3.20\,GPa respectively. In the first four of these cases no negative energy phonon modes were observed, indicating that the computed structure remained dynamically stable. For the calculation with a lattice parameter of 9.16\,\AA\ a large number of negative energy modes were observed, indicating that for such a compressed lattice the structure becomes unstable. From this point onward when we refer to these DFT calculations we mean those with $9.21 \leq a \leq 9.23$\,\AA.

When comparing the computed dispersion with that measured experimentally, an empirical scale factor was applied to the eigenvalues (energies) but not the eigenvectors (structure factors). This was determined by finding the simulation that agrees most closely with the data after such a uniform re-scale: the structure factors from the simulations with different lattice parameters were fitted to the neutron data with a single adjustable parameter (the energy scale). For \zwo the best overall agreement was found for the case of $a = 9.215$\,\AA\,\, and a scale factor $\sim 0.95$. This approach was used because changing the lattice parameter for the simulation does not affect the eigenvalues uniformly.

Additional calculations were performed with the same DFT parameters and a lattice parameter of 9.215\,\AA, but using the PBEsol functional \cite{PBESol-2008}, which has been reported as yielding values for the bulk properties (elastic constants) of \zwo\ closer to the experimentally determined values than other functionals \cite{WECK-2018}.

\subsection*{C: Machine learned potentials and molecular dynamics calculations}\label{subsec:MLMD}

Calculations were performed using the automated potential development (APD) workflow \cite{duff2023automated}, which is open-source and available from gitlab \cite{apd-webpage}. APD automates all DFT calculations, potential optimizations and potential simulations, as well as any subsequent cycles of active-learning that may be required to refine the potential. In addition APD also generates various properties at the DFT and potential level. Note that the DFT calculations performed by APD are distinct from those used elsewhere.

CASTEP with optimized norm-conserving pseudo-potentials \cite{Pseudopot-Ramer-1999,rappe1990optimized} was used as the DFT engine for APD.  Calculations were performed using two generalized gradient approximation exchange-correlation functionals, both PBE \cite{Perdew_PRL_1996_Generalized} and the PBEsol  functional \cite{perdew2008restoring}. For all except the phonon calculations a Monkhorst–Pack grid of 2$\times$2$\times$2 (with 44 atom cell) was used. For configuration generation, a relatively low cutoff of 520\,eV was used. Molecular dynamics (MD) simulations were run at 333\,K, 666\,K and 1000\,K, each for 2000 configurations with a timestep of 7\,fs. Uncorrelated snapshots (separated by $\sim$ 15 configurations) were then recalculated using a cutoff of 1000\,eV, ensuring an accuracy of $\pm 1$\,meV/atom in energies. This cutoff, and the k-points grid, were automatically optimized by APD to achieve this accuracy.

Moment tensor potentials (MTPs) \cite{shapeev2016moment} were then optimized using the MLIP2 software package \cite{novikov2020mlip}, and fitted to energies, forces and stresses (EFS) from the DFT simulations. Five distinct potentials (MTP1-5) were independently optimized. Once optimized, MTP1-5 were used to drive LAMMPS (Large-scale Atomic/Molecular Massively Parallel Simulator) molecular dynamics simulations \cite{plimpton2007lammps}. LAMMPS jobs were performed up to 1000 K in 100 K increments, each for 10 ps using the Nos\'e-Hoover thermostat and barostat. From these simulations, configurations for which EFS predictions are unreliable were identified using MLIP's MaxVol algorithm, using an extrapolation grade of 3 as a threshold. These configurations were then recalculated using DFT, and subsequently used to reoptimize potentials. The procedure was repeated until no configurations exceed an extrapolation grade of 3 during the LAMMPS simulations conducted at all temperatures. This is handled separately (and automatically) for each potential, so that each undergoes its own active-learning cycle.

As an additional post-processing step, dynamic structure factors were computed for a selection of q-points and across a range of temperatures. Dynamical structure factors were calculated using the Dynasor open-source framework \cite{fransson2021dynasor}. Calculations were performed at high-symmetry points $Y=(0, 0.5, 0)$, $M = (0.5, 0.5, 0)$, and the midpoint between these points. A total of 200 runs, each lasting 1\,ns, were performed for each temperature 100\,K, 200\,K, 300\,K and 400\,K. Each run included a 200\,ps equilibration with the Number Volume Temperature (NVT) ensemble, followed by a 60\,ps equilibration and a 400\,ps production run with the Number Temperature Energy (NTE) ensemble.  Structure factors were calculated from each run and subsequently averaged.

\section*{Appendix II: Distinguishing phonon modes via their structure factors}

The coherent inelastic neutron scattering cross-section from single-phonon excitations (creation) is given by:

\begin{eqnarray}
\left( \frac{d^{2}\sigma}{d\Omega dE_{f}} \right) &=& \frac{k_{i}}{k_{f}} \frac{\sigma_{\rm{coh}}}{4\pi} \frac{(2\pi)^{3}}{v_{0}} \frac{1}{2M} e^{-2W}\times\nonumber\\
&& \sum_{s} \sum_{\mathbf{G}}  \frac{( \mathbf{Q} \cdot \mathbf{e}_{s} )^{2}}{\omega_{s}} \langle n_{s} +1 \rangle \times\nonumber\\
&& \delta(\omega - \omega_{s})\delta(\mathbf{Q}-\mathbf{q}-\mathbf{G})
\label{eq:cross-sec}
\end{eqnarray}

\noindent where $k_{i}$ and $k_{f}$ are the incident and final neutron wavevectors, $\sigma_{\rm{coh}}$ is the coherent neutron cross-section of the material being studied, $v_{0}$ is the sample volume, $M$ the mass of the sample, $W$ the Debye-Waller factor, $\mathbf{Q} = \mathbf{G}+\mathbf{q}$ the excitation wavevector, composed of the Brillouin zone vector $\mathbf{G}$ and the reduced phonon wavevector $\mathbf{q}$, $\mathbf{e}_{s}$ is the phonon eigenvector for mode $s$ with frequency $\omega_{s}$, and $\langle n_{s}+1\rangle$ is the Bose-Einstein population factor. The key term is the one involving the dot product of the wavevector $\mathbf{Q}$ and the phonon eigenvector $\mathbf{e}_{s}$, which implies that the scattering cross-section is largest when $\mathbf{Q}$ and $\mathbf{e}_{s}$ are parallel, and zero when they are perpendicular. By measuring with many different orientations of $\mathbf{Q}$ all of the phonons will give rise to a signal in the INS measurement. Such an approach permits modes which may be similar in energy but with different eigenvectors to be resolved.

We note that single crystal inelastic neutron scattering measurements of \zwo have been reported \cite{Mittal2007}, but with data collected at only a few high symmetry points in a small number of Brillouin zones, and without sufficient energy resolution to discern subtle shifts with a high degree of certainty. Inelastic neutron scattering experiments with single crystal samples have been performed on the structurally simple NTE materials ReO$_{3}$ \cite{Axe-PRB-1985} and Cu$_{2}$O \cite{Beg-PRB-1976} in the 1970s and 1980s, again covering only a few high symmetry points in a small number of Brillouin zones. More recently, as time-of-flight neutron spectrometers have developed, the multi-Brillouin zone measurement technique used in this study has been employed on structurally simple compounds exhibiting anomalous thermal expansion, e.g. Cu$_{2}$O \cite{Saunders-PRB-2022} and NaBr \cite{Shen-PRL-2020}, permitting a detailed analysis of the phonon spectra.

\section*{Appendix III: Calculated phonon dispersion and density of states}

\begin{figure}
\centering
    \includegraphics[scale=0.43]{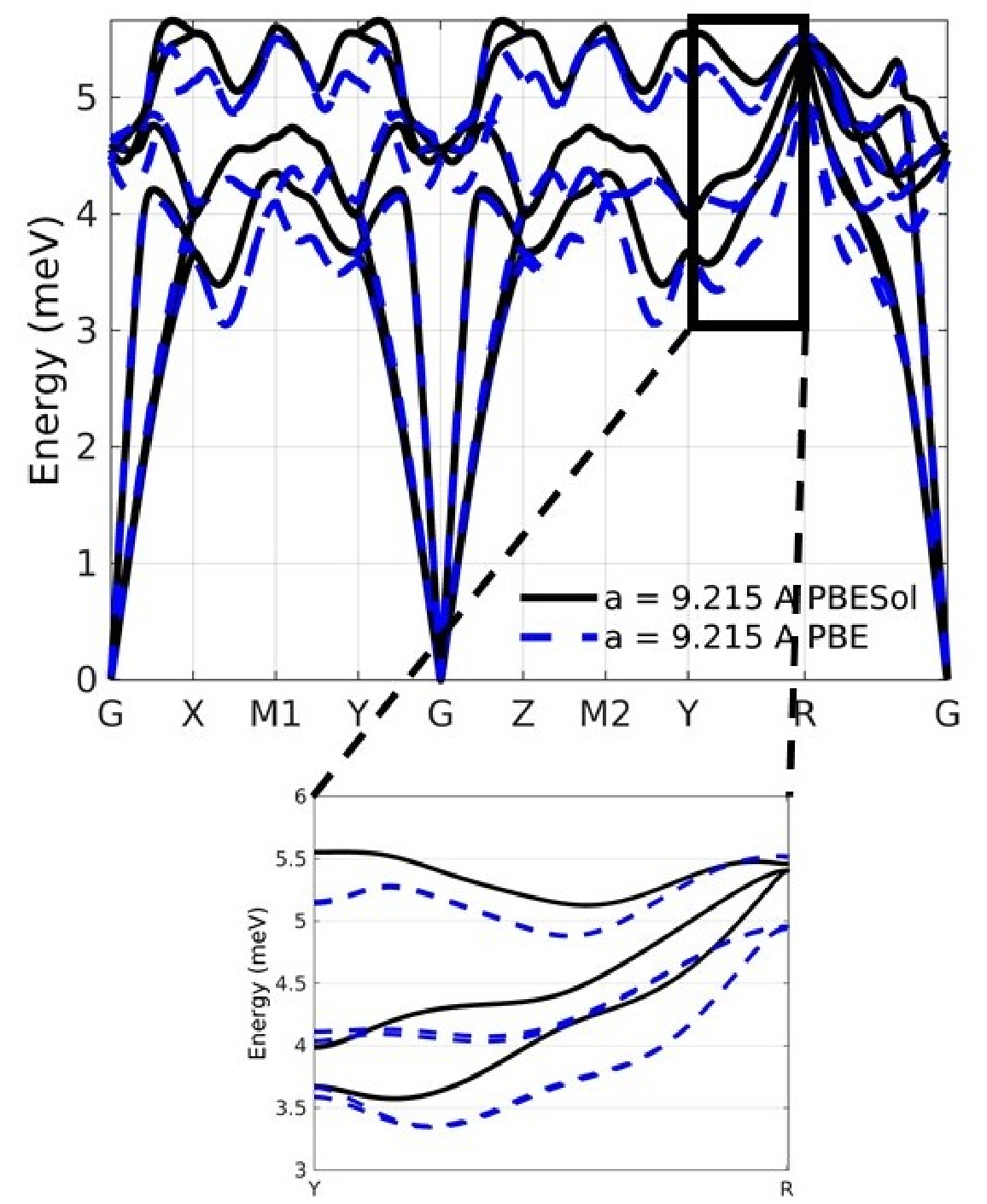}
    \caption{Comparison of dispersion from DFT calculations with different choices of functional, with the same lattice parameter of $a = 9.215$\,\AA. The reciprocal space path is the same as described in fig. 2 of the main text. The solid black curve is for calculations with the PBEsol functional and the dashed blue curve is for calculations with the PBE functional (the same as the dashed blue curve in fig. 7 of the main text). The inset shows a zoomed in section of the dispersion between 3 and 6\,meV.}
    \label{f:disp_func}
\end{figure}

In figure \ref{f:disp_func} we compare calculations for the PBE and PBEsol functionals with the same lattice parameter of 9.215\,\AA. It can be seen that using the results from the PBEsol functional result in slightly higher phonon energies for the same lattice parameter. As with the comparison between different lattice parameters using the PBE functional, shown in the main text, we note that at certain trajectories and points there are more obvious differences. For example, the inset to fig. \ref{f:disp_func} shows that the modes at the R point are significantly closer together in energy for the PBEsol functional than they are for PBE.

We therefore conclude that for analysing the INS data, calculations with either the PBE or the PBEsol functional are valid. On this basis we chose not to proceed with further simulations using the PBEsol functional (e.g. full geometry and lattice parameter optimization) since there would likely be little to be gained over the PBE calculations already performed.

Qualitatively, it would seem that the modes which are most sensitive to perturbations to the choice of psuedopotential are also the modes that seem to be most difficult for any of the DFT calculations to reproduce, i.e. those modes between about 3 and 5\,meV. These are also the modes with the largest magnitude mode Gr\"uneisen parameters and hence are likely to be those most sensitive to anharmonic effects.

\begin{figure}
\centering
    \includegraphics[scale=0.5]{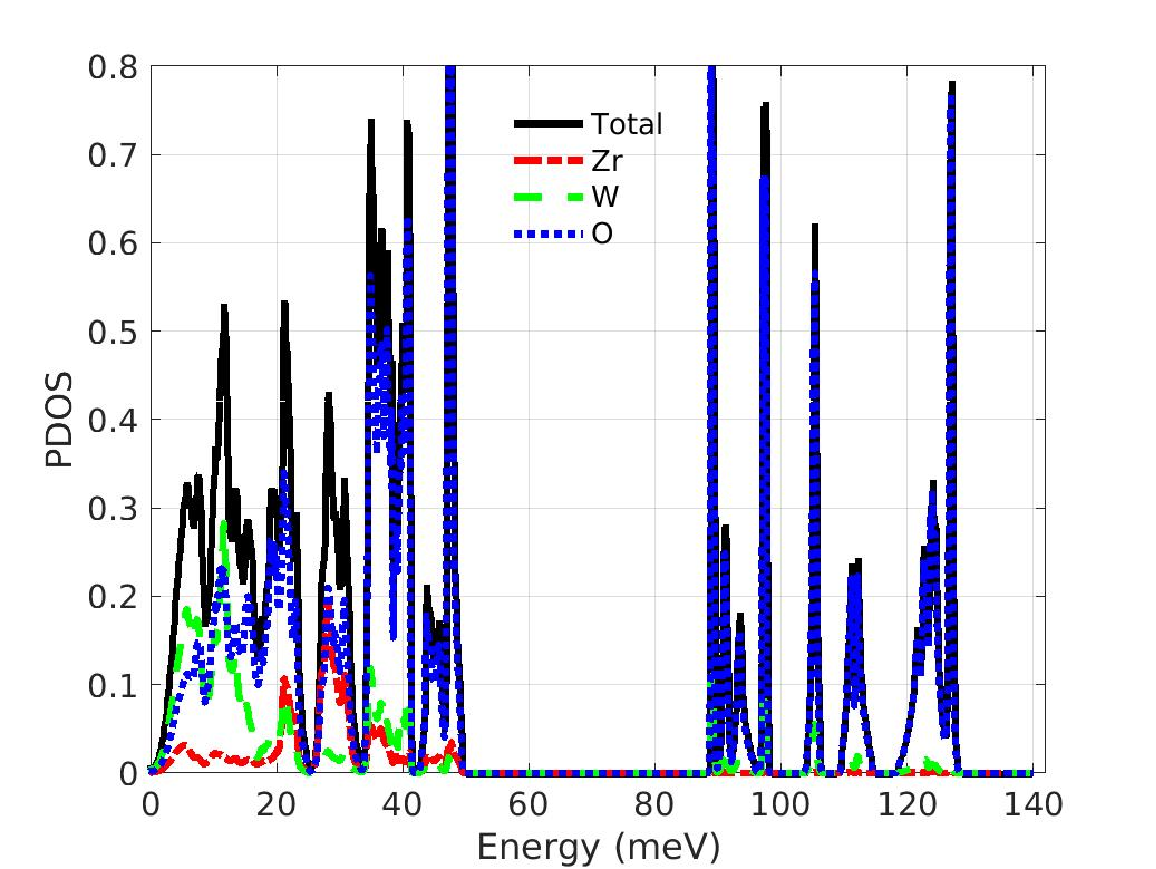}
    \caption{The phonon density of states (PDOS) as a function of energy computed using DFT. The total PDOS is shown by the black curve, and the element-wise contributions are shown in red (zirconium), green (tungsten) and blue (oxygen).}
    \label{f:PDOS_wide}
\end{figure}

Next, we consider the phonon density of states (PDOS). In fig. \ref{f:PDOS_wide} we show the partial phonon density of states (PDOS) computed from DFT, broken down into the contributions from each of the constituent atomic species. The advantage of computing this property is that it is intrinsically averaged across all of reciprocal space and hence offers some insight into the combined effect of many different phonon modes to the material’s thermal properties. The disadvantage is that the averaging means that information about particular phonon modes is lost.

In the energy range probed by our INS experiment, the PDOS has approximately equal weight from tungsten and oxygen, but very little contribution from zirconium. This indicates that the zirconium ions are much more fixed in position, somewhat characteristic of rigid-unit-modes. Further decomposition of the PDOS down to the individual atom level (not shown) confirms that the contribution from individual tungsten ions is larger than for individual oxygen ions by about a factor two to three at low energies, but the oxygen ions are more numerous so sum up to give a comparable contribution to the PDOS.

\begin{figure}
\centering
    \includegraphics[scale=0.5]{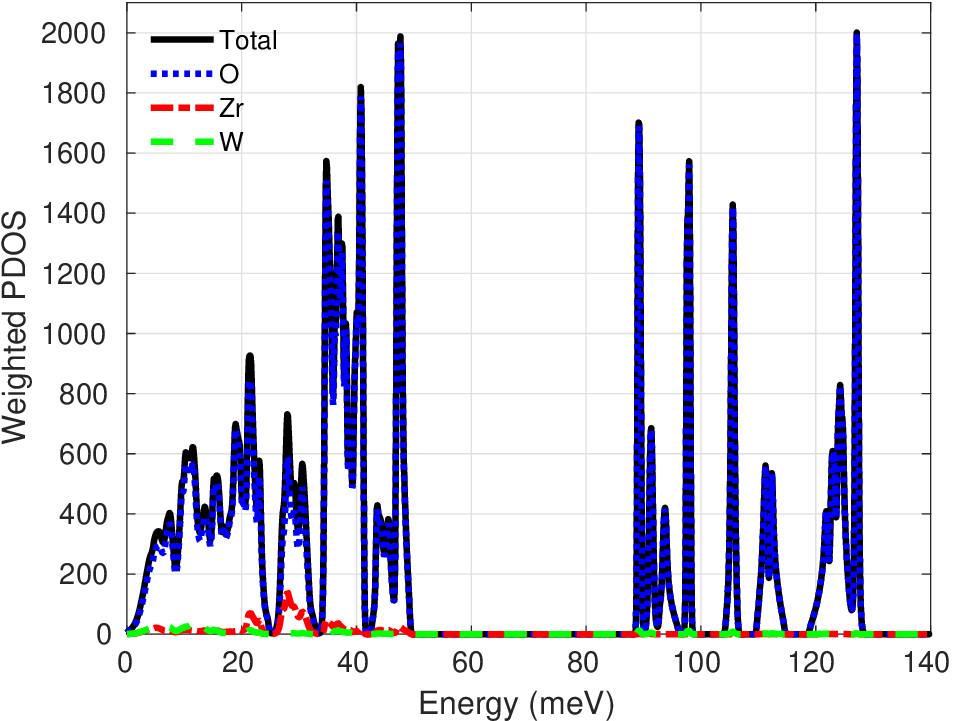}
    \caption{The phonon density of states (PDOS) weighted by the element-specific neutron scattering factors, as a function of energy computed using DFT. The total PDOS is shown by the black curve, and the element-wise contributions are shown in red (zirconium), green (tungsten) and blue (oxygen).}
    \label{f:PDOS_weighted}
\end{figure}

Fig. \ref{f:PDOS_weighted} shows the PDOS with neutron weighting applied (see eq. \ref{eq:cross-sec}), which allows direct comparison with the neutron structure factor measured in an experiment, such as that in ref.\,\onlinecite{Ernst1998}. The scattering lengths of all three constituent elements are similar in magnitude, as are their Debye-Waller factors, so the largest effect on the neutron cross-section comes from the division of all the other terms by the atomic mass. This gives significantly more weight to the oxygen modes than to tungsten and zirconium, with the consequence that the neutron scattering signal is dominated by oxygen motion. 

\begin{figure*}
\centering
    \includegraphics[scale=0.9]{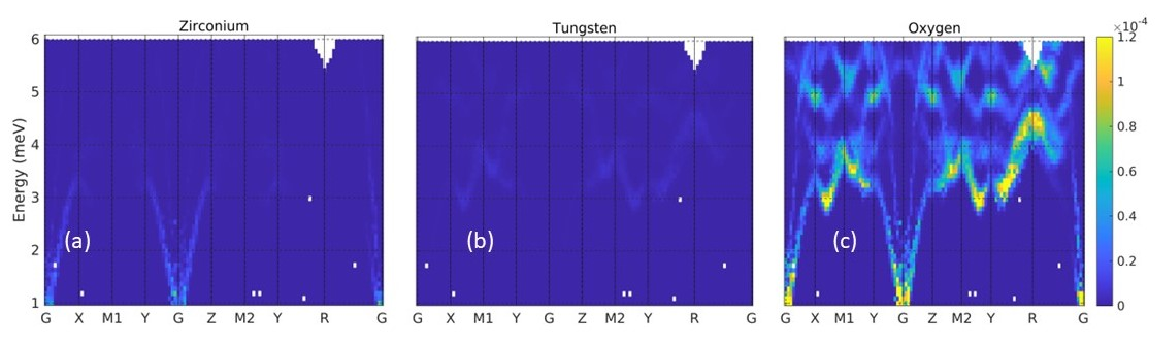}
    \caption{Slices through the $\mathbf{Q}$ - energy plane, following the same scheme as fig. 2 of the main text. The colormaps show the neutron scattering intensity calculated using DFT, decomposed into the contributions from each individual element. Panel (a) shows the contribution from zirconium, panel (b) the contribution from tungsten, and panel (c) the contribution from oxygen. Note that the intensity scale is the same as that in fig. 3 of the main text and is the same for all three panels.}
    \label{f:Element_spag}
\end{figure*}

We can decompose the simulations shown in figs. 2 -- 4 of the main text, into the contributions from the different atomic species. Examples are shown in fig. \ref{f:Element_spag}, for the $(-2,2,2)$ Brillouin zone center, i.e. equivalent to fig. 3 from the main text. Most of the signal arises from oxygen motions, with contributions from tungsten and zirconium around an order of magnitude smaller. A similar effect is seen for computation of the phonon density of states. However, although the neutron measurements are less directly sensitive to the motion of the heavier metal atoms, the fact that the DFT gives good agreement for the oxygen means that we can be confident that its predictions for the zirconium and tungsten motion are accurate, since each of them are surrounded by a cage of oxygen ions.

\section*{Appendix IV: Structure factors from MD calculations}

Here we show the neutron structure factor $S(\mathbf{q},\omega)$ computed from the MD calculations at several fixed values of $\mathbf{q}$, for temperatures of 100, 200, 300 and 400\,K. The peaks were fitted using Gaussians, and it is the center positions of these that are plotted in fig. 10 of the main text. As well as the peaks moving to higher energies with increasing temperature, one can also see that, as would be expected, they also broaden.

\begin{figure}
\centering
    \includegraphics[scale=0.6]{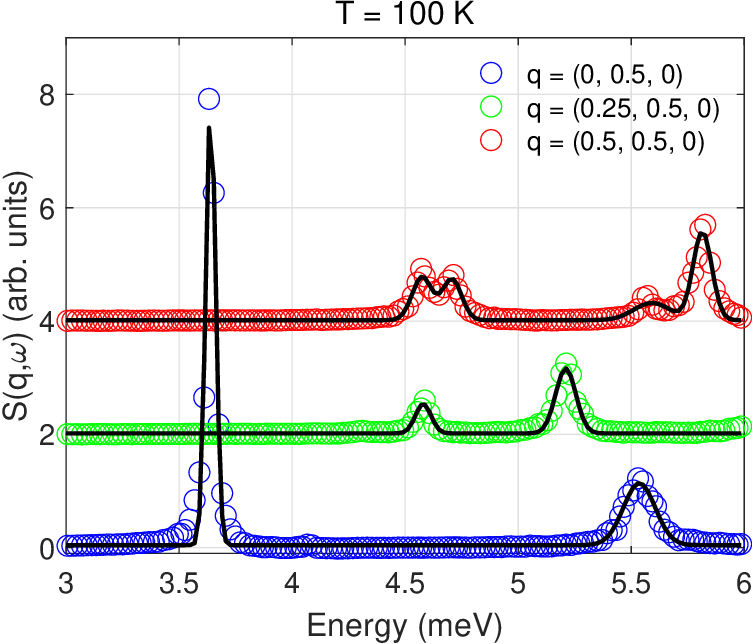}
    \caption{Energy scans of the MD-computed neutron structure factor $S(\mathbf{q},\omega)$ at fixed wavevectors $\mathbf{q} = (0, 0.5, 0)$ (blue circles) , $\mathbf{q} = (0.25, 0.5, 0)$ (green circles), and $\mathbf{q} = (0.5, 0.5, 0)$ (red circles). Data are successively offset on the y-scale by 2 units. Fits of the peaks to a Gaussian lineshape are shown by black lines. These data are from the MD simulation at $ T = 100$\,K.}
    \label{f:MD_Escan_100}
\end{figure}

\begin{figure}
\centering
    \includegraphics[scale=0.6]{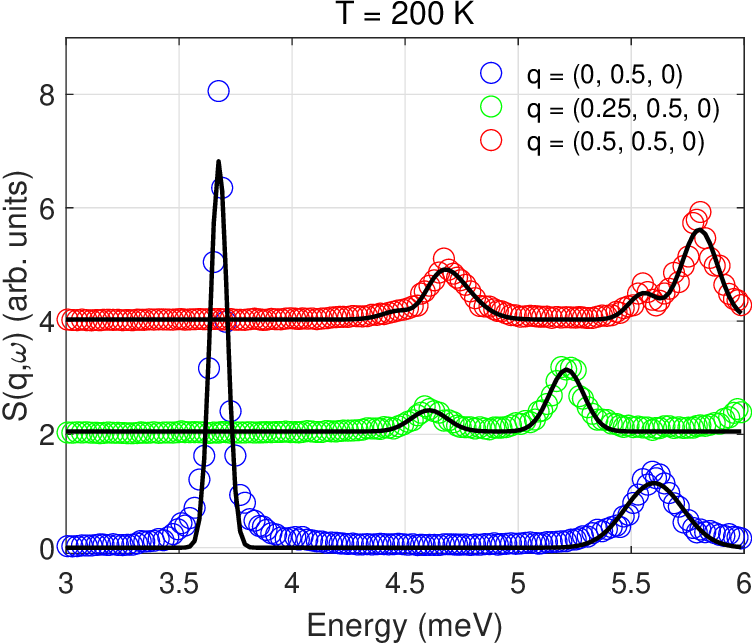}
    \caption{Energy scans of the MD-computed neutron structure factor $S(\mathbf{q},\omega)$ at fixed wavevectors $\mathbf{q} = (0, 0.5, 0)$ (blue circles) , $\mathbf{q} = (0.25, 0.5, 0)$ (green circles), and $\mathbf{q} = (0.5, 0.5, 0)$ (red circles). Data are successively offset on the y-scale by 2 units. Fits of the peaks to a Gaussian lineshape are shown by black lines. These data are from the MD simulation at $ T = 200$\,K.}
    \label{f:MD_Escan_200}
\end{figure}

\begin{figure}
\centering
    \includegraphics[scale=0.6]{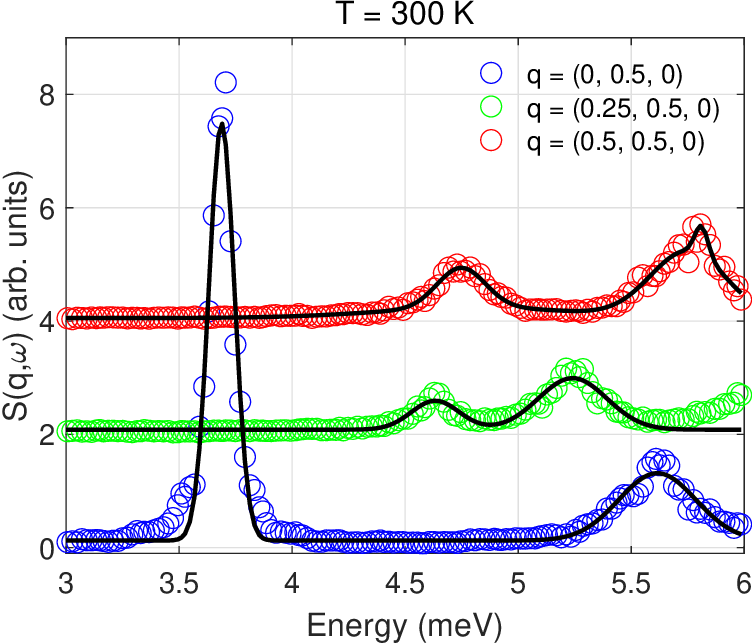}
    \caption{Energy scans of the MD-computed neutron structure factor $S(\mathbf{q},\omega)$ at fixed wavevectors $\mathbf{q} = (0, 0.5, 0)$ (blue circles) , $\mathbf{q} = (0.25, 0.5, 0)$ (green circles), and $\mathbf{q} = (0.5, 0.5, 0)$ (red circles). Data are successively offset on the y-scale by 2 units. Fits of the peaks to a Gaussian lineshape are shown by black lines. These data are from the MD simulation at $ T = 300$\,K.}
    \label{f:MD_Escan_300}
\end{figure}

\begin{figure}
\centering
    \includegraphics[scale=0.6]{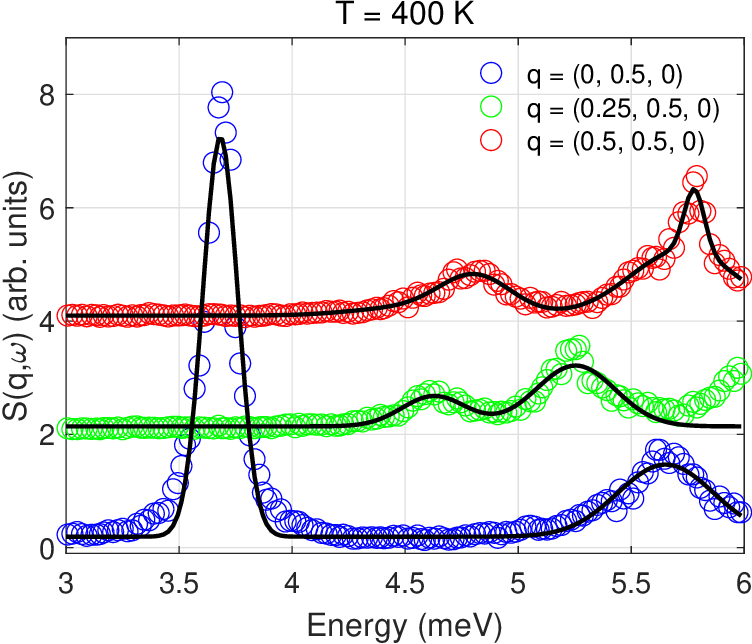}
    \caption{Energy scans of the MD-computed neutron structure factor $S(\mathbf{q},\omega)$ at fixed wavevectors $\mathbf{q} = (0, 0.5, 0)$ (blue circles) , $\mathbf{q} = (0.25, 0.5, 0)$ (green circles), and $\mathbf{q} = (0.5, 0.5, 0)$ (red circles). Data are successively offset on the y-scale by 2 units. Fits of the peaks to a Gaussian lineshape are shown by black lines. These data are from the MD simulation at $ T = 400$\,K.}
    \label{f:MD_Escan400}
\end{figure}

\clearpage

\twocolumngrid

\bibliography{ZWO_references}

\end{document}